# Sulfur Ice Astrochemistry: A Review of Laboratory Studies

**Authors and Affiliations**

Duncan V. Mifsud[1,†], Zuzana Kaňuchová[2], Péter Herczku[3], Sergio Ioppolo[4], Zoltán Juhász[3], Sándor T. S. Kovács[3], Nigel J. Mason[1], Robert W. McCullough[5], Béla Sulik[3]

1: Centre for Astrophysics and Planetary Science, School of Physical Sciences, University of Kent, Canterbury CT2 7NH, United Kingdom

2: Astronomical Institute, Slovak Academy of Sciences, Tatranská Lomnica SK-059 60, Slovak Republic

3: Institute for Nuclear Research (ATOMKI), Debrecen H-4001, PO Box 51, Hungary

4: School of Electronic Engineering and Computer Science, Queen Mary University of London, London E1 4NS, United Kingdom

5: Department of Physics and Astronomy, School of Mathematics and Physics, Queen's University Belfast, Belfast BT7 1NN, United Kingdom

† Corresponding Author: duncanvmifsud@gmail.com

**ORCID Numbers**

| | |
|---|---|
| Duncan V. Mifsud | 0000-0002-0379-354X |
| Zuzana Kaňuchová | 0000-0001-8845-6202 |
| Péter Herczku | 0000-0002-1046-1375 |
| Sergio Ioppolo | 0000-0002-2271-1781 |
| Zoltán Juhász | 0000-0003-3612-0437 |
| Sándor T. S. Kovács | 0000-0001-5332-3901 |
| Nigel J. Mason | 0000-0002-4468-8324 |
| Béla Sulik | 0000-0001-8088-5766 |

**Word Count**

13,753 words (excluding title page, references, and figures/captions)

**Acknowledgements**

Duncan V. Mifsud is the grateful recipient of a University of Kent Vice-Chancellor's Research Scholarship which has allowed him to pursue his doctoral studies. The research of Zuzana Kaňuchová is supported by VEGA – the Slovak Grant Agency for Science (Grant No. 2/0023/18) and by the Slovak Research and Development Agency (Contract No. APVV-19-0072). Sergio Ioppolo gratefully acknowledges the Royal Society for financial support. The authors all gratefully acknowledge support from the Europlanet 2024 RI which has received funding from the European Union's Horizon 2020 research innovation programme under Grant Agreement No. 871149. The authors are also especially grateful to all the technical and operational staff at the Tandetron Laboratory, ATOMKI Institute for Nuclear Research in Debrecen (Hungary) for their continued support and assistance.

# Sulfur Ice Astrochemistry: A Review of Laboratory Studies

## Abstract

Sulfur is the tenth most abundant element in the universe and is known to play a significant role in biological systems. Accordingly, in recent years there has been increased interest in the role of sulfur in astrochemical reactions and planetary geology and geochemistry. Among the many avenues of research currently being explored is the laboratory processing of astrophysical ice analogues. Such research involves the synthesis of an ice of specific morphology and chemical composition at temperatures and pressures relevant to a selected astrophysical setting (such as the interstellar medium or the surfaces of icy moons). Subsequent processing of the ice under conditions that simulate the selected astrophysical setting commonly involves radiolysis, photolysis, thermal processing, neutral-neutral fragment chemistry, or any combination of these, and has been the subject of several studies. The *in-situ* changes in ice morphology and chemistry occurring during such processing has been monitored via spectroscopic or spectrometric techniques. In this paper, we have reviewed the results of laboratory investigations concerned with sulfur chemistry in several astrophysical ice analogues. Specifically, we review (i) the spectroscopy of sulfur-containing astrochemical molecules in the condensed phase, (ii) atom and radical addition reactions, (iii) the thermal processing of sulfur-bearing ices, (iv) photochemical experiments, (v) the non-reactive charged particle radiolysis of sulfur-bearing ices, and (vi) sulfur ion bombardment of and implantation in ice analogues. Potential future studies in the field of solid phase sulfur astrochemistry are also discussed in the context of forthcoming space missions, such as the NASA James Webb Space Telescope and the ESA Jupiter Icy Moons Explorer mission.

## Keywords

Sulfur; Astrochemistry; Laboratory Astrophysics; Review; Radiolysis; Photolysis

## 1    Introduction

Sulfur is the tenth most abundant element in the universe. Atomic sulfur has an abundance of $1.32 \times 10^{-5}$ relative to hydrogen (Asplund et al. 2009), while the unipositively charged ion has a relative abundance of $1.66 \times 10^{-5}$ (Esteban et al. 2004). The most common isotope, $^{32}S$, accounts for ~95% of all sulfur in the universe and is produced via silicon burning within stars at temperatures of $> 2.5 \times 10^9$ K (McSween and Huss 2010). This nuclear reaction forms part of the so-called *alpha ladder* which produces elements in abundance, thus explaining the natural relative ubiquity of sulfur. Nuclear fusion of oxygen atoms may also account for the formation of $^{32}S$.

$$^{28}\text{Si} + {}^{4}\text{He} \rightarrow {}^{32}\text{S} + \gamma$$

$$^{16}\text{O} + {}^{16}\text{O} \rightarrow {}^{32}\text{S} + \gamma$$

Chemically speaking, sulfur is located within the same periodic group as oxygen (both elements are chalcogens; Group XVI) and is thus predicted to have a similar chemistry. However, sulfur chemistry is in fact more versatile: the lower electronegativity of sulfur compared to oxygen allows sulfur to accommodate a positive charge more readily, while still permitting the formation of hydrogen bonds. Additionally, sulfur is capable of commonly exhibiting multiple oxidation states between –2 and +6, while oxygen is more limited in this regard.

Sulfur is also known to exist as multiple allotropes. Indeed, sulfur is second only to carbon in terms of the number of known allotropic forms (Greenwood and Earnshaw 1997). These allotropes largely take the form of chains and rings, with the most common being the puckered ring structure of $S_8$. In addition to the known allotropes, several polymorphs are also known to exist. The most common structures are the α- (rhombic) and β- (monoclinic) polymorphs of $S_8$.

From a biochemical point of view, sulfur (along with hydrogen, carbon, oxygen, nitrogen, and phosphorus) is considered to be one of the six elements fundamental to life. It is found in a wide variety of biomolecules, such as amino acids, nucleic acids, sugars, and vitamins. The distribution of sulfur throughout the cosmos, however, is certainly not uniform, and varying amounts of different sulfur-bearing compounds have been detected in the interstellar and intergalactic media, on planetary surfaces, in atmospheres and exospheres, on icy moons, and in cometary material.

The known occurrence of sulfur in various astrophysical environments, coupled with its pivotal role in biological systems therefore makes it an element of particular interest with respect to studies in astrochemistry and astrobiology. Accordingly, the space chemistry of this element has become an increasingly popular topic of study, a summary of which follows below.

## 1.1 Sulfur in the Interstellar Medium

Although the idea of active chemistry in interstellar space was initially considered impossible due to low densities and temperatures, evidence for the presence of molecules arose in 1937 (Swings and Rosenfeld 1937) and was later confirmed by McKellar (1940) through his detection of CH and CN radicals. This was soon followed by the detection of $CH^+$ by Douglas and Herzberg (1941). Such results showed that chemistry within interstellar and circumstellar environments was indeed possible, and prompted further research into the exact mechanisms and processes by which molecules form.

Since these early findings, the fields of astrophysical chemistry and molecular astrophysics have grown into highly interdisciplinary fields, drawing on expertise from physical chemistry, atomic and molecular physics, astronomy, planetary science, and geochemistry, among others (van Dishoeck 2014; 2017; Jørgensen et al. 2020). Today, we are aware of the existence of about 200 molecules in interstellar and circumstellar regions, as well as of around 70 extragalactic molecules (for a complete list, see the referenced Cologne Database for Molecular Spectroscopy).

Of the currently confirmed detected interstellar molecules, only 23 contain sulfur atoms. In the case of extragalactic molecules, this number drops to 12. This apparent lack of chemical diversity in astrophysical sulfur-bearing molecules is somewhat reflective of a greater problem in astrochemistry: that while within diffuse regions and more primitive interstellar environments, gas phase sulfur satisfactorily accounts for its total cosmic abundance (Sofia et al. 1994; Jenkins 2009), within dense molecular clouds and star-forming regions there is an unexpected paucity of molecular sulfur (Anderson et al. 2013). In such dense regions the observed gas phase summed abundances of SO, $SO_2$, $H_2S$, and CS constitute only a fraction of the expected amount (Tieftrunk et al. 1994).

A similar trend is encountered within the solid phase, where OCS and $SO_2$ are the dominant interstellar molecules (Palumbo et al. 1995; 1997; Boogert et al. 1997; Ferrante et al. 2008). Their abundances, however, only account for < 5% of the total expected sulfur abundance (Boogert et al. 1997; Palumbo et al. 1997). The need to account for this so-called *depleted sulfur* has led some to suggest that it may be locked within hitherto undetected reservoirs, such as within icy grain mantles or as refractory material (Smith 1991; van der Tak et al. 2003). However, some studies have questioned whether refractory sulfurous materials can make up for this missing sulfur (Woods et al. 2015). As such, the question of sulfur depletion continues to be a pressing issue in contemporary astrochemistry (Wakelam et al. 2011; Jiménez-Escobar and Muñoz-Caro 2011; Laas and Caselli 2019).

## 1.2  Sulfur in Solar and Planetary Systems

Sulfur is also prevalent within solar and planetary systems such as our own. The presence of solid phase or gas phase sulfur (or both) has been established, for instance, on each of the inner terrestrial planets. Examples include: solid phase elemental sulfur within the permanently shadowed polar regions of Mercury (Sprague et al. 1995), sulfur oxides in the atmosphere of Venus (Vandaele et al. 2017a; 2017b), sulfate minerals found across the surface of Mars (King and McLennan 2010), and within practically every environmental reservoir on Earth, including the atmosphere, geosphere, hydrosphere, and biosphere (Brimblecombe 2013).

Perhaps one of the most interesting areas of research within planetary science, particularly as it pertains to sulfur ice astrochemistry, is the case of the Galilean moons of Jupiter. These moons are (in increasing orbital radius to their host planet) Io, Europa, Ganymede, and Callisto (Fig. 1). The rotations of all these moons are tidally locked to Jupiter, meaning that the same side continually faces the host planet.

Io, the innermost and the only non-icy Galilean moon, is the most geologically active celestial body in the Solar System, having at least 400 volcanoes at its surface (Lopes et al. 2004). Volcanic processes on this moon result in the emission of $SO_2$ and other sulfur-bearing molecules. These ejected molecules are ionised within the Jovian magnetosphere, which is known to contain a wealth of energetic ions and plasmas from other sources such as the solar wind and sputtered surface ices originating from the other moons (Delitsky and Lane 1998).

The icy surfaces of Europa, Ganymede, and Callisto are subject to continuous bombardment by Jovian magnetospheric ions, which can be implanted into the ice and may even engender chemical reactivity. From a sulfur chemistry perspective, this is interesting as it involves the implantation and reaction of sulfur ions in surface ices, or the bombardment of surface sulfur ices by other projectiles. Given the theorised presence of subsurface liquid oceans on these moons, the products of such chemical reactions and processes are of great interest to the astrobiology community. The global flux of incoming energetic charged projectiles for each of the Galilean moons, as given by Johnson et al. (2004) is shown in Table 1.

**Table 1:** Global energy fluxes of charged particles at the surfaces of the Galilean moons.

| Galilean Satellite | Global Average Energy Flux (keV s$^{-1}$ cm$^{-2}$) |
|---|---|
| Io | $1 \times 10^9$ |
| Europa | $5\text{-}8 \times 10^{10}$ |
| Ganymede | $2\text{-}50 \times 10^8$ |
| Callisto | $2 \times 10^8$ |

The chemical and energetic compositions of the magnetospheric charged particles bombarding the icy moons has also been studied using data collected by the NASA Galileo mission (Paranicas et al. 2002; Mauk et al. 2004). Briefly summarised, a plethora of species can be found in the Jovian magnetosphere, but at distances of $> 7$ $R_J$ (where $R_J$ is taken to be the radius of Jupiter at 71,492 km), sulfur and oxygen ions dominate the energetic ($> 50$ keV) ion density while at distances of 20-25 $R_J$ protons dominate both the integral number and energy densities (Mauk et al. 2004). Indicative energy spectra for protons, oxygen ions, and sulfur ions near the orbits of Europa and Ganymede are provided below (Fig. 2).

The sulfur radiochemistry of the Galilean moons is thus of astrochemical and astrobiological significance, particularly in light of the presence of liquid oceans beneath their icy surfaces. As such, this astrophysical setting is often simulated during laboratory investigations so as to elucidate radiolytic products and likely reaction mechanisms.

### 1.3  The Need for Laboratory Experiments

Laboratory astrochemistry studies offer planetary and space scientists the opportunity to simulate the conditions at a particular astrophysical setting (such as the interstellar medium or planetary surfaces) and investigate chemical reactions induced via some processing technique. In the case of sulfur ice astrochemistry, this generally involves the formation of an ice of known chemical composition which is deposited onto a cold (5-200 K) substrate transparent to certain wavelengths of light, thus allowing for spectroscopic monitoring of ice composition and morphology. The ice is then processed thermally, photochemically, radiochemically, or by some other means and reaction products are deduced. A fuller explanation of the range of experimental techniques used is available in the review by Allodi et al. (2013).

The data generated from such laboratory investigations is necessary in the interpretation of data collected by both past and future space missions. Two upcoming missions which may be of particular interest to sulfur astrochemistry are the NASA James Webb Space Telescope (JWST) and the ESA Jupiter Icy Moons Explorer (JUICE) mission (Gardner et al. 2006; Grasset et al. 2013). JWST will allow infrared (IR) observations of both gas phase and solid phase sulfur molecules in several extra-terrestrial environments, including interstellar and extragalactic space, outer Solar System objects (including the Galilean moons of Jupiter), and exoplanetary atmospheres. JUICE will focus on a comparative characterisation of Europa, Ganymede, and Callisto and will delve into compositional mapping of their surfaces. In the case of Europa, there will be a real focus on ice chemistry and organic molecules relevant to prebiotic chemistry.

It must be noted, however, that contemporary laboratory astrochemistry experiments are largely non-systematic, often reporting the results of very specific temperature, pressure, and processing conditions. Although such experiments contribute to our scientific understanding of likely astrochemical reactions, it does raise some questions as to how widely applicable the results of such studies are. For instance, a study looking into the photolytic or radiolytic processing of an ice at 20 K is applicable to the interstellar medium, but perhaps not so to the surfaces of the Galilean moons where temperatures are often much higher.

In reality, there is a need for systematic studies in which a single experimental factor (e.g. ice chemical composition) is held constant while other factors (e.g. ice morphology, temperature, processing type, processing energy, etc.) are varied. Such studies will provide more detailed information on the dependence of reaction products on such parameters, and will be useful in assessing the applicability of previous experiments to different astrophysical settings. Additionally, there is also a need to increase characterisation of reaction kinetics and yield analysis so as to determine whether or not reaction products may accumulate in the studied astrophysical setting over geological or astronomical time-scales.

## 1.4 An Overview of this Review

This paper will concern itself with reviewing the results of laboratory investigations in the field of sulfur ice astrochemistry. Though it is important to mention that sulfur astrochemistry is also an explored theme in mineralogy, isotope geochemistry, cosmochemistry, and gas phase studies, these subjects will not be tackled here. Furthermore, a complete review of the laboratory techniques employed during experiments on sulfur ice astrochemistry goes beyond the scope of this review, although an extensive summary was provided by Allodi et al. (2013).

The motivation behind this review is to serve as a repository for solid phase experiments in sulfur astrochemistry. A thorough understanding of this subject is important, particularly in light of the forthcoming JWST and JUICE missions, as data from laboratory experiments would assist these missions in addressing some major questions in sulfur space chemistry, such as the depleted sulfur problem in dense interstellar clouds and the potential of the Galilean moons to host some form of life.

This review describes and summarises the results of studies in six parts according to experiment type: (i) spectroscopic studies of candidate solid phase interstellar sulfur molecules, (ii) atom (or radical) addition reactions, (iii) thermally-induced chemistry, (iv) photochemical processing, (v) radiolytic processing using chemically inert charged projectiles, and (vi) reactive sulfur ion bombardment and implantation. The ordering of these subjects follows an energy gradient, starting at one end with atom and radical addition reactions which have no activation energy requirement and finishing at the other with sulfur ion radiolysis which can involve high energy (> 1 MeV) projectiles.

The reader will appreciate that there may be some overlap between spectroscopic and photochemical studies (given in Sections 2 and 5, respectively). However, an effort has been made such that each section may be viewed as a stand-alone review of a particular aspect of sulfur ice astrochemistry.

## 2    Spectroscopy of Candidate Sulfur-Bearing Molecules

Of the 23 sulfur-bearing molecules known to exist in interstellar and circumstellar settings, many have thiol (S–H), thioketone (C=S), and sulfinyl (S=O) functional groups. This provides valuable information when deciding on which species are likely candidates for future detection in interstellar space. Positive detections of molecules in interstellar environments have largely been made via radio astronomy; a technique which requires the target molecule to possess a sufficiently large dipole moment. As such, several molecules which lack this property cannot be detected by radio astronomy, and spectroscopic methods are more appropriate in this regard.

However, for successful detections of interstellar molecules to be made spectroscopically, it is necessary to have their corresponding laboratory-generated spectra at relevant temperatures. For example, recent studies have explored in great detail the IR and ultraviolet (UV) spectra of thiol compounds, including methanethiol, ethanethiol, 1-propanethiol, and 2-propanethiol (Hudson 2016; 2017; Pavithraa et al. 2017a; 2017b; Hudson and Gerakines 2018). Although detections of the two smaller molecules in interstellar environments have been confirmed (Linke et al. 1979; Kolesniková et al. 2014), the latter two remain candidate molecules (Gorai et al. 2017).

These studies have provided additional information related to different solid phases encountered at low temperatures as well as the influence of conformational isomer effects. For instance, methanethiol exists as an amorphous ice at temperatures < 65 K. However, upon heating to just above this temperature, crystallisation occurs and two phases are produced (Hudson and Gerakines 2018): a thermodynamic product (α-phase) and a kinetic product (β-phase). These phases produce slightly different IR spectra (Fig. 3), so knowledge of their absorbance peaks would aid in the identification of this molecule when observed in an astrophysical setting and in the determination of which phase is present.

In the case of ethanethiol, Pavithraa et al. (2017a) showed that warming of a low-temperature ice resulted in a phase change from amorphous to crystalline at 110 K. Interestingly, further warming of the ice caused the phase to switch back to amorphous at 125 K. These phase changes were found to be reversible, also occurring during cooling of the ice.

When the length of an aliphatic carbon chain increases sufficiently (as in the case of 1-propanethiol), there exists the possibility that molecules of the same species may differ in structure only by rotation around a C–C bond at very low temperatures. These individual structures are referred to as *conformational isomers* or *rotamers*. As an amorphous ice containing such isomers is warmed, less stable isomers rotate so as to adopt a more stable structure. This re-orientation results in changes in the corresponding IR spectra. Such spectral changes have been reported for 1-propanethiol (Hayashi et al. 1966; Torgrimsen and Klæboe 1970).

Computational analysis has also been used in deciphering the spectral signatures of molecules of astrophysical relevance. To continue using methanethiol as an example: microwave, IR, and UV spectra have been extensively studied by modelling rotational, vibrational, and electronic excitement of this molecule (May and Pace 1968; Schlegel et al. 1977; Mouflih et al. 1988; Zakharenko et al. 2019; etc.). Furthermore, computational studies have also been used to understand conformational isomerism in thiols and thioesters (Fausto et al. 1987).

Aside from the abovementioned thiols, other candidate molecules containing sulfur atoms have been studied using spectroscopic techniques. These molecules include ethenethiol, ethynethiol, ethanethial, ethenethione, thiirane, and thiirene. They are of particular interest either because their oxygen analogues have already been identified in interstellar environments, or because they produce highly abundant molecular fragments (Lee et al. 2019; Martin-Drumel et al. 2019; McGuire et al. 2019).

Further spectroscopic studies of a variety of sulfur-bearing molecular species will thus increase our spectral repository, and will assist space-based telescopes in positive identifications of new extra-terrestrial molecules. However, future studies should emphasise the importance of mid-IR characterisations, especially in light of the forthcoming JWST mission ($\lambda = 0.6\text{-}28.3$ μm) and the recently retired Spitzer Space Telescope ($\lambda = 3.6\text{-}160$ μm), both of which included the mid-IR in their operational spectra.

## 3     Neutral-Neutral Addition Experiments

Neutral-neutral additions refer to the process of bond formation between a target molecule and another neutral species, generally an atom or radical. Such additions do not require the input of energy to overcome an activation energy barrier, and so can occur efficiently at very low temperatures (~10 K). These reactions are most evident within the interiors of dense molecular clouds in the interstellar medium, which are cold (10-20 K) and dark as they are shielded from impinging light or radiation (Linnartz et al. 2015). As such, thermochemistry, photochemistry, and radiochemistry are disfavoured compared to neutral-neutral additions.

In their review, Linnartz et al. (2015) highlighted the findings of some laboratory atom addition experiments (particularly hydrogenation reactions), as well as three processes by which these reactions are thought to occur: the Langmuir-Hinshelwood, Eley-Rideal, and Harris-Kasemo (or *hot atom*) mechanisms. Briefly explained, the Langmuir-Hinshelwood mechanism involves two atoms adsorbing onto an ice grain, reaching thermal equilibrium, and then migrating towards each other and reacting.

The Eley-Rideal mechanism involves an atom impacting an already-adsorbed atom and reaction occurring before thermal equilibrium can take place. Finally, the Harris-Kasemo mechanism involves an atom adsorbing onto an ice grain and migrating towards a second atom which is in thermal equilibrium and subsequently reacting with that atom before it itself can reach equilibrium.

It appears that, although sulfur neutral-neutral reactions in the gas phase have been studied rigorously, analogous solid phase experiments are extremely scarce. As the focus of this paper is the solid phase, we will not review the results of gas phase experiments, but rather we direct the interested reader to a series of papers on this subject, the first three of which are cited herein

(Strausz and Gunning 1962; Knight et al. 1963a; 1963b). Other papers (Becker et al. 1974; Prasad and Huntress 1980; Kaiser et al. 1999; etc.) are also available.

When it comes to solid phase neutral-neutral reactions involving sulfur, the exiguity of laboratory-based studies means that, in order to provide an adequately thorough overview of the subject, our review must also extend to modelling experiments. To avoid any ambiguity, the results of computational work will be labelled as such in the following review. The recent review by Cuppen et al. (2017) provides a good overview on the treatment of astrochemical reactions, including neutral-neutral reactions, on dust grains by computational simulations.

Computational assessments by Laas and Caselli (2019) showed that sulfur atom additions on grains at interstellar temperatures can result in catenation reactions to produce $S_n$ ($n$ = 2, 3, 4), which in turn can partake in diradical ring-closure reactions to form cyclic $S_m$ ($m$ = 5, 6, 7, 8). Although the activation energy barriers for such reactions are essentially zero, the binding energies to the grain for the product allotropes are assumed to increase with the number of sulfur atoms involved. As such, the heavier allotropes have a higher binding energy which does not permit efficient thermal roaming and so they can be destroyed by alternative methods before any further reaction can take place.

In their experimental investigation, Jiménez-Escobar and Muñoz-Caro (2011) discussed the formation of $S_8$ as a result of cryogenic elongation reactions mediated by the HS radical (Barnes et al. 1974). Addition of two HS radicals to each other was proposed to explain the observed presence of $H_2S_2$. Results from modelling have shown that this formation mechanism is possible (Zhou et al. 2008), but unlikely as it would need to compete with the hydrogenation of HS to $H_2S$, which is known to be highly efficient. Instead, Zhou et al. (2008) proposed that sulfur atom addition to $H_2S$ could yield $H_2S_2$ or its isomer $H_2SS$, however such a reaction is probably inefficient due to the high binding energies of the reactants.

Deeyamulla and Husain (2006) showed that a neutral-neutral insertion reaction between atomic carbon and $H_2S$ could result in the formation of HCS and a hydrogen atom. A further hydrogenation reaction between the products would result in the formation of $H_2CS$, although this reaction is limited by the amount of HCS within the ice available for reaction. Atomic carbon is also known to react with HS radicals to yield CS, which is abundant in both the gas and solid phases in interstellar space. However, within the solid phase, the presence of CS is also thought to be dependent upon the direct accretion of gas phase CS onto dust grains.

Solid phase CS adsorbed onto icy interstellar dust grains may go on to react with CH to form $C_2S$ and a hydrogen atom, the reaction benefitting from the high abundance of each reactant (Kaiser 2002). Longer carbon-sulfur chains can also be formed through neutral-neutral reactions, with modelling work by Laas and Caselli (2019) showing that carbon atom addition to $C_2S$ results in $C_3S$, while sulfur atom addition to $C_4H$ yields $C_4S$ and a hydrogen atom.

The sulfur-analogue of methanoic acid, dithiomethanoic acid, may also be produced via solid phase neutral-neutral reactions. In fact, its synthesis in the solid phase has been computationally predicted to be more efficient than that for the gas phase (Laas and Caselli). This process involves the combination of CS and HS radicals to yield CSSH, which can then be hydrogenated to give the final acid product.

The neutral-neutral combination formation mechanism for methanethiol is interesting since the solid phase formation route for the analogous alcohol, methanol, has been extremely well-described (Kaiser 2002; Linnartz et al. 2015). The formation of methanethiol within astrophysical ices is thought to be similar, proceeding through barrierless radical-radical reactions between H and either $CH_2SH$ or $CH_3S$ (Gorai et al. 2017). An alternative formation mechanism involving the hydrogenation of $H_2CS$ through a series of high activation energy reactions has also been proposed (Vidal et al. 2017).

Other thiol molecules have also been investigated: Gorai et al. (2017) and Gorai (2018) studied the solid phase neutral-neutral reactions that lead up to the formation of ethanethiol, 1-propanethiol, and 2-propanethiol via computational means. These reactions were observed to often involve the addition of S, H, $CH_3$, and (less commonly) $C_2H_5$ fragments and were previously considered in the modelling and observational work of Hasegawa and Herbst (1993) and Müller et al. (2016).

The addition of a nitrogen atom to HS results in the formation of NS and a hydrogen atom. This reaction is interesting for two reasons: firstly, it is thought to be an important sink for solid phase sulfur, accounting for up to 10% of the total sulfur budget (Laas and Caselli 2019). Secondly, polymeric compounds comprised of sulfur-nitrogen bonds at very low temperatures have been noted to display a wide variety of useful and interesting properties, such as superconductivity and metallicity (Chivers 2005). The ubiquity of NS in space environments, however, somewhat contrasts the sparsity of laboratory studies on nitrogen-sulfur bond formation in astrophysical ice analogues.

With respect to SO, which is more commonly found in the icy mantles of dust grains than in the gas phase, this molecule may be formed via direct combination of a sulfur atom and an oxygen atom, or alternatively via the reaction of a sulfur atom with OH or via the reaction between an oxygen atom and SH (Laas and Caselli 2019). SO may be oxygenated to yield $SO_2$, however this reaction is known to be inefficient if SO reacts with an oxygen atom due to the high binding energy of the latter species. Instead, this reaction is more likely to occur via the reaction between SO and $O_2$. The addition of excited atomic oxygen ($^1D$) to $SO_2$ was observed to yield $SO_3$ in the laboratory work by Schriver-Mazzuoli et al. (2003).

OCS can also be formed through atom addition reactions in solid ices, mainly through the reaction of sulfur atoms with CO (Laas and Caselli 2019). Although primarily destroyed via photo-dissociation reactions, OCS can also be destroyed by scavenging sulfur atoms which react with it to form $S_2$ and CO. It is important to note that such a formation mechanism is applicable only to the solid phase, as gas phase formation of OCS via atom addition usually occurs as a result of the addition of an oxygen atom to HCS or the addition of a sulfur atom to HCO.

As can be seen, neutral-neutral sulfur chemistry in astrophysical settings is potentially rich and varied, yet laboratory-based experiments in this field are hard to come by. Future research should thus attempt to fill this gap in knowledge, especially since a better understanding of low temperature and pressure sulfur atom (or radical) reactions could go some way in helping to answer the question of depleted sulfur in dense interstellar clouds as the products of such reactions are often refractory. Furthermore, additional experimental data would be invaluable to the modelling community as it would allow for the results of computational simulations to be tested against empirical evidence.

# 4    Thermal Processing

Thermal studies in astrophysical ice analogues constitute an important aspect of research. For example, thermal desorption studies are key to understanding the chemical availability of species which partake in star and planet formation, as well as in the synthesis of prebiotic molecules. A review of such studies was provided by Burke and Brown (2010). Aside from such desorption studies, thermally-induced chemistry has also been investigated as such reactions are known to occur in various astrophysical environments where temperatures are high enough to overcome the relevant activation energy barriers (for a review, see Theulé et al. 2013). The results of those experiments which are pertinent to sulfur ice astrochemistry will be discussed in this section.

Kaňuchová et al. (2017) investigated the thermochemistry of both pure $SO_2$ and $SO_2$:$H_2O$ mixed ices in the temperature range 16-160 K. Their findings showed that pure $SO_2$ ice begins to sublimate very efficiently at 120 K. Like Collings et al. (2004) and Jiménez-Escobar and Muñoz-Caro (2011) before them, Kaňuchová et al. (2017) also noticed that $SO_2$ sublimates at a higher temperature when included in a $H_2O$ matrix.

When studying $SO_2$:$H_2O$ ice mixtures, it was noted that thermochemical reactions took place and the main reaction product was $HSO_3^-$, with smaller amounts of $S_2O_5^{2-}$ also being produced (Kaňuchová et al. 2017). These results complemented the previous findings of Loeffler and Hudson (2010; 2013; 2015), who also showed that in the presence of $H_2O_2$ (which is the main irradiation product of $H_2O$ ice and thus of significance to many astrophysical contexts), $HSO_3^-$ goes on to form $HSO_4^-$ and $H_3O^+$. Subsequent deprotonation of $HSO_4^-$ by $H_2O$ yields $SO_4^{2-}$ and $H_3O^+$ ions.

$$H_2O + SO_2 \rightarrow H^+ + HSO_3^-$$
$$H^+ + HSO_3^- + H_2O_2 \rightarrow H_3O^+ + HSO_4^-$$
$$HSO_4^- + H_2O \rightarrow H_3O^+ + SO_4^{2-}$$
$$2\,HSO_3^- \rightarrow H_2O + S_2O_5^{2-}$$

Such reactions are of importance in the context of two Galilean moons of Jupiter: Europa and Ganymede. The surfaces of both these worlds are dominated by $H_2O$ ice, and $SO_2$ has been detected on both (this will be discussed in more detail in Sections 6 and 7). Therefore, the above reactions constitute a non-radiolytic and non-photolytic formation mechanism for $SO_4^{2-}$, $HSO_4^-$, $HSO_3^-$, and $S_2O_5^{2-}$ on the surfaces of these icy moons.

Recent work by Bang et al. (2017) has also shown that thermally-driven reactions between $H_2O$ ice and gas phase $SO_2$ are possible. Their study showed that $SO_2$ molecules adsorbed at the surface of a $H_2O$ ice can react with the ice at temperatures of > 90 K. The primary reaction products are $SO_2^-$, $HSO_3^-$, and $OH^-$. Programmed heating of the physisorbed gas to 120 K causes desorption from the ice surface and thus separates it from the chemisorbed hydrolysis products. Quantum chemical calculations suggest that the mechanism of formation of these products is similar to the equations outlined above (Bang et al. 2017).

Building on their previous work in which they showed that thermally-driven reactions occur in mixed ices of $H_2O$:$H_2O_2$:$SO_2$, Loeffler and Hudson (2016) went on to show that such reactions are also possible if the oxidant is changed from $H_2O_2$ to $O_3$, which is also a radiolysis product of $H_2O$. The results of this study showed that $O_3$ ice is consumed producing $HSO_4^-$, although the reaction sequence begins with a reaction between $SO_2$ and $H_2O$ which is not too dissimilar to that which occurs in the Earth's atmosphere (Erickson et al. 1977; Penkett et al. 1979).

$$H_2O + SO_2 \rightarrow H^+ + HSO_3^-$$
$$2\ HSO_3^- \rightarrow H_2O + S_2O_5^{2-}$$
$$HSO_3^- + O_3 \rightarrow HSO_4^- + O_2$$

Given that the experiments of Loeffler and Hudson (2016) were conducted in the temperature range 50-120 K, they are of direct relevance to the Galilean satellites Europa, Ganymede, and Callisto, where mean surface temperatures are ~100 K and where sulfur-bearing ices are mixed with $H_2O$ and $O_3$ (or its precursor $O_2$). Detections of $O_3$ have been made on the trailing side of Ganymede (Noll et al. 1997). This is to be expected, as this side contains surface $O_2$ and is subject to preferential bombardment by Jovian magnetospheric ions.

$SO_2$ has also been detected on the Galilean moons, although its distribution is somewhat more elusive (McCord et al. 1998a). The results of the study by Loeffler and Hudson (2016) suggest that $SO_2$ should be depleted on the trailing side of Ganymede if mixed in ices along with both $H_2O$ and $O_3$. However, given that the consumption of $O_3$ is dependent on a prior reaction between $SO_2$ and $H_2O$ (as outlined in the reaction sequence above), it is possible that pure frosts of $SO_2$ may co-exist alongside pure $O_3$ ices.

In the case of Europa, $SO_2$ has been detected on the trailing hemisphere, but seems to be absent on the leading hemisphere (Hendrix and Johnson 2008; Hendrix et al. 2011). The fact that $O_2$ is present on both hemispheres should lead one to believe that $O_3$ should only be detected on the leading side. However, to date, no detections of this molecule have been made on the surface of Europa, meaning that there may be some unknown reaction consuming $O_3$ in this hemisphere.

The case of Callisto is somewhat more difficult to interpret. Although $O_2$ has been detected in the trailing hemisphere (Spencer and Calvin 2002), no $O_3$ has been identified. $SO_2$ has only been detected in the leading hemisphere, which is unexpected as this side is less susceptible to magnetospheric ion bombardment. The identification of $SO_2$ on Callisto has, however, been challenged, with carbonate species being suggested as an alternative for the spectral observation (Johnson et al. 2004).

There is also the potential for these results to be applied to cometary chemistry. The Rosetta mission identified the presence of $SO_2$ in the coma of comet of 67P/Churyumov-Gerasimenko, but did not detect $O_3$ (Bieler et al. 2015). It is possible that the formation of sulfur oxyanions via the reaction of cometary $SO_2$ and $H_2O$ ices are responsible for the absence of $O_3$.

Other sulfur-bearing molecules have also been the focus of research with regards to thermally-driven reactions in astrophysical environments. The thermal reaction between oxygen atoms and $CS_2$ to produce OCS is one example (Ward et al. 2012). Mahjoub and Hodyss (2018) investigated the reaction between OCS and methylamine over temperatures of 12-300 K. These molecules have been identified in comets, which are known to experience thermal processing as they orbit around the sun. Their results showed that, at temperatures exceeding 100 K, methylammonium methylthiocarbamate is formed as a result of the nucleophilic attack of methylamine on OCS (Fig. 5). This product molecule may be an intermediate in the formation of peptides, and thus relevant to studies in astrobiology and prebiotic chemistry.

# 5 Photochemical Processing

Photochemical reactions are among the most important in astrochemistry and planetary science, and their investigation has enabled atmospheric chemistry and dynamics both within the interstellar medium and many planetary systems to be quantified. Several studies and reviews on sulfur chemistry within the atmospheres of Earth, Solar System planets and moons, and exoplanets are available (Sze and Ko 1980; Moses et al. 1995; 2002; Colman and Trogler 1997; Zahnle et al. 2009; Tian et al. 2010; Whitehall and Ono 2012; Hu et al. 2013; Hickson et al. 2014; Ono 2017; He et al. 2020; etc.).

Although of great importance to our understanding of planetary science, atmospheric studies largely deal with gas phase chemistry. In this review, we are more concerned with the solid phase and so this section will limit itself to the discussion of sulfur ice photochemistry and the results of laboratory investigations.

## 5.1 The Importance of Spectroscopic Studies in Photochemistry

In Section 2, reference was made to the fact that detailed knowledge of the absorption spectra of candidate interstellar molecules would aid in their search and identification. Molecular photochemistry usually begins via the absorption of a photon which results in excitation of the absorbing species to higher electronic energy states, or in the dissociation of the molecule through the process of bond fission (Wells 1972).

As such, characterisation of the absorption spectra of astrochemically relevant molecules is required because it provides a starting point for understanding their photochemical reactivity. This is especially true in the case of extreme- and vacuum-UV photons, as well as X-rays, which initiate much of the photochemistry in interstellar environments and planetary surfaces (Pilling and Bergantini 2015; Öberg 2016).

As a brief example, the vacuum-UV absorption spectrum of $SO_2$ ice has been studied in some detail (Holtom et al. 2006; Mason et al. 2006) and distinct spectral signatures have been detected which allow for phase discrimination between amorphous and crystalline ices. These features may be used to glean further information as to the structure of the ice as a function of substrate temperature and rate of deposition: for example, rapid deposition rates at low temperatures are amenable to the formation of amorphous ices (Mason et al. 2006).

This has important astrochemical implications, as deposition time-scales on dust grains in the interstellar medium are likely to be much longer than those which can be reproduced in the laboratory. Thus, there is good reason to suggest that $SO_2$ ice in the interstellar medium is in fact crystalline, or at least a mixture of crystalline and amorphous ice. Reflectance spectra of astrophysical $SO_2$ do not show any evidence of crystallinity (Nash et al. 1980; Hapke et al. 1981), however they only include few data points and limited resolution may not allow for the detection of such crystalline features.

Given that radiolysis induced by X-ray absorption can occur in interstellar and circumstellar environments, such as near T-Tauri phase stars (Gullikson and Henke 1989; Pilling and Bergantini 2015; Pilling 2017), there is also some scope for recording the X-ray absorbance spectra of astrophysical ice analogues. X-ray absorption studies have been carried out in the past in order to determine the speciation of sulfur in coal and petroleum (Spiro et al. 1984; George and Gorbaty 1989; Huffmann et al. 1991; Waldo et al. 1991), soils (Morra et al. 1997; Xia et al. 1999), microbial biochemical products (Pickering et al. 1998; 2001; Prange et al. 2002), and batteries (Cuisinier et al. 2013; Pascal et al. 2014).

These studies largely made use of either X-ray absorption near edge structure (XANES) spectroscopy or X-ray absorption fine structure (XAFS) spectroscopy. Despite the established use of X-ray spectroscopy in other fields, to the best of the authors' knowledge such a technique has yet to be used to study laboratory generated astrophysical sulfur ice analogues and so there is some potential for future studies in this regard. A review on the use of XANES in the determination of sulfur oxidation states and functionality in complex molecules was provided by Vairavamurthy (1998).

## 5.2  Laboratory Photochemistry Experiments

With respect to investigating the photochemical reactivity of sulfur-bearing ice analogues, $SO_2$ and $H_2S$ ices have received the most attention (Cassidy et al. 2010). UV photolysis of solid phase $SO_2$ was examined in detail by Schriver-Mazzuoli et al. (2003) who considered the irradiation of pure $SO_2$ ice, as well as $SO_2$ ice trapped in an excess of amorphous $H_2O$ ice, with photons of $\lambda = 156$, 165, and 193 nm.

Results showed that photolysis of the pure ice resulted in the formation of $SO_3$, while the major photolysis product of the $SO_2$:$H_2O$ ice mixture was $H_2SO_4$. In the case of pure $SO_2$ ice photolysis, there is also some evidence to support the photo-reaction occurring after the formation of $SO_2$ dimers (Sodeau and Lee 1980).

Photolysis of $SO_2$:$H_2O$ ice mixtures using far-UV photons of $\lambda = 147$, 206, 254, and 284 nm has been recently performed and results showed that the main photolysis products were the sulfur oxyanions $SO_4^{2-}$, $HSO_4^-$, and $HSO_3^-$ (Hodyss et al. 2019). Interestingly, although photons with $\lambda > 219$ nm are not energetic enough to cause the dissociation of $SO_2$, these products were also observed for both the $\lambda = 254$ and 284 nm photolysis experiments. These observations were explained through the reaction of electronically excited $SO_2$ reacting with a ground-state $SO_2$ molecule (Hodyss et al. 2019). This reaction mechanism is thought to be similar to that for the gas phase where the reaction products are SO and $SO_3$ (Chung et al. 1975), which may then go on to react with $H_2O$ to produce the sulfur oxyanions observed. Alternatively, the excited-state $SO_2$ molecule may react directly with ground-state $H_2O$.

$$SO_2\,(^3B_1) + SO_2\,(X,\,^1A_1) \rightarrow SO + SO_3$$

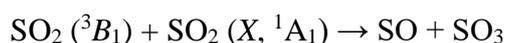

These studies are of great importance in the context of planetary science, particularly in the case of the icy Galilean satellites, upon the surfaces of which $SO_2$ has been detected (Lane et al. 1981). Hence, photolytic reactions represent a method of $SO_2$ depletion and $H_2SO_4$ production at the surfaces of these moons. Although there is good evidence to suggest that the major formation mechanism of $H_2SO_4$ is via magnetospheric sulfur ion implantation (discussed in more detail in Section 7), these photochemical results are nonetheless important and may be extended to other icy Solar System bodies.

Experiments have also been carried out in order to determine the chemical effect of soft X-rays (< 2 keV) on $SO_2$ ices. Such experiments are entirely appropriate in the context of interstellar astrochemistry, as $SO_2$ ices have been detected near young stellar objects where X-ray intensities are higher (Boogert et al. 1997). Laboratory results showed that $SO_3$ is the product of such radiolysis (de Souza Bonfim et al. 2017), with the major formation mechanisms shown below:

$$SO + O_2 \rightarrow SO_3$$

$$SO_2 + O \rightarrow SO_3$$

$$SO_2 + O^+ \rightarrow SO_3^+ + e^- \rightarrow SO_3$$

Soft X-ray radiolysis experiments of $SO_2$ mixed ices have also been performed. Irradiation of $H_2O:CO_2:NH_3:SO_2$ ices at 50 K and 90 K resulted in the formation of $SO_3$, $H_2SO_4$, and associated sulfur oxyanions (Pilling and Bergantini 2015). Interestingly, $OCN^-$ was also detected among the radiolysis products, and is likely to be formed as a result of the dissociation of $CO_2$ and $NH_3$ and the recombination of the resultant fragments. However, radiolytic dissociation and fragment recombination did not result in sulfur bonding to any new elements.

The photolysis of $H_2S$ ices represents another major aspect of sulfur astrochemistry. Comprehensive work in this regard was performed by Jiménez-Escobar and Muñoz-Caro (2011), who irradiated $H_2S$ ice with UV light at 7 K. The results of this experiment showed that $H_2S$ photolysis is fairly rapid and produces a large variety of species including HS, $H_2S_2$, $HS_2$, and $S_2$. This work complemented previous findings which showed that photolysis of $H_2S$ adsorbed on LiF at 110 K resulted in the dissociation to H and HS radicals and the formation of $H_2$ gas (Harrison et al. 1988; Liu et al. 1999; Cook et al. 2001).

The UV irradiation of $H_2S$ ice has most recently been revisited by Zhou et al. (2020), who wished to address the apparent depletion of $HS/H_2S$ ratios observed in the interstellar medium relative to those predicted by contemporary models. Their experimental results revealed a wavelength dependence for the quantum yield of HS formation via $H_2S$ photo-dissociation. Also taking into account $H_2S$ parent molecule absorption and the interstellar radiation field, such a result implies that just ~26% of interstellar photo-excitations result in the successful production of HS radicals. It is therefore necessary to reconsider some of the outcomes of computational sulfur astrochemistry studies.

When in a matrix of $H_2O$ ice, the UV irradiation of $H_2S$ yielded products such as $SO_2$, $SO_4^{2-}$, $HSO_3^-$, $HSO_4^-$, $H_2SO_2$, $H_2SO_4$, and $H_2S_2$ (Jiménez-Escobar and Muñoz-Caro 2011). An interesting observation made during this study was that, although the sublimation temperature of pure $H_2S$ was noted to be 82 K this value rose significantly when mixed in a matrix of $H_2O$ molecules. The reason for this is the trapping of $H_2S$ molecules by less volatile $H_2O$ molecules in an analogous fashion to clathrates. Thus, $H_2S$ only co-sublimates at higher temperatures with $H_2O$. Such results are in agreement with previous findings (Collings et al. 2004).

Photo-irradiation of $H_2S$ ices mixed with other species have also been studied. Chen et al. (2015) irradiated $H_2S:CO$ and $H_2S:CO_2$ ices at 14 K with UV and extreme-UV photons and determined the nature of the sulfur-bearing product molecules. They found that both ice mixtures produce OCS, with the formation efficiencies being greater for the $H_2S:CO$ mixture and when lower starting concentrations of $H_2S$ were used. Other sulfur-bearing molecules were detected among the photolysis products: $CS_2$ was produced after the photo-processing of the $H_2S$:CO mixture, and $SO_2$ was produced after the photo-processing of the $H_2S$:$CO_2$ ice.

An interesting study of vacuum-UV photolysis of gas phase $H_2S$ mixed with either ethene or 1,3-butadiene found that the main products are refractory thiol compounds (Kasparek et al. 2016). Though not an investigation into sulfur ice chemistry, this result carries interesting implications for interstellar sulfur chemistry, as it reveals a method of locking sulfur-bearing material away as refractories and thus aids in our understanding of the sulfur depletion problem discussed in Section 1 (Jenkins 2009; Laas and Caselli 2019).

The photo-processing of more complex $H_2S$ mixed ices has also been performed. UV irradiation of a $H_2O$:CO:$NH_3$:$H_2S$ ice resulted in a vast array of products such as polymeric chains of sulfur, nitrogen-based heterocycles, and cyclic-( S–$CH_2$–NH–$CH_2$–NH–$CH_2$), while irradiations of $H_2S$:CO and $H_2S$:$CH_3OH$ mixed ices produced $H_2S_2$, $HS_2$, $CS_2$, $H_2CO$, and OCS on the one hand, and CO, $CO_2$, $CH_4$, $H_2CO$, and $CS_2$ on the other (Jiménez-Escobar and Muñoz-Caro 2011; Jiménez-Escobar et al. 2014).

Aside from $H_2S$ and $SO_2$, investigations in other simple astrochemically relevant molecules, namely OCS and $CS_2$, have been carried out. Photo-irradiation of monolayers of these species adsorbed on both amorphous and polycrystalline $H_2O$ ices were carried out by Ikeda et al. (2008) at 90 K, who showed that the photolysis of these species by UV photons of $\lambda$ = 193 nm led to the formation of $S_2$ via the combination of atomic sulfur with OCS or $CS_2$, or alternatively via the combination of dissociated sulfur atoms. The work of Ikeda et al. (2008) built upon previous efforts by Leggett et al. (1990) and Dixon-Warren et al. (1990), who also observed $S_2$ formation when irradiating OCS adsorbed to LiF at 166 K using $\lambda$ = 222 nm photons.

Cryogenic photochemistry between OCS and halogen species has also been investigated: the reactions with $Cl_2$, $Br_2$, and BrI have been performed in a solid Ar matrix at 15 K using broadband UV photons (Romano et al. 2001; Tobón et al. 2006). Various reaction products were formed, the most chemically interesting of which were *syn*-halogenocarbonylsulfenyl halides based on X–C(=O)–S–Y backbones, where X and Y are the constituent atoms of the dihalide originally incorporated into the ice (Fig. 6). The corresponding *anti*-rotamer products were not detected, and are known to be less stable (Romano et al. 2001; Tobón et al. 2006).

When similar experiments were used to investigate the photochemistry between $CS_2$ and the dihalides $Cl_2$, $Br_2$, and ClBr in a Ar ice matrix, both *syn*- and *anti*-halogenothiocarbonylsulfenyl halides based on X–C(=S)–S–Y backbones were detected (X and Y once again being the constituent atoms of the original dihalide) among other products (Fig. 6; Tobón et al. 2007).

Solid phase photochemistry of more complex, exotic molecules may yield significant insights into the chemistry of different functional groups found in the interstellar medium. For instance, Zapała et al. (2019) recently examined the photolysis of 2-sulfanylethanenitrile in a Ar matrix at 6 K. A related compound, sulfanylmethanenitrile, has already been detected in the interstellar medium (Halfen et al. 2009). Their results showed that, among some photo-dissociation products formed via the loss of –CN and –SH groups, several isocyano compounds were produced as a result of photo-isomerisation processes (Zapała et al. 2019; Fig. 7).

Pharr et al. (2012) investigated the individual, solid phase 10 K UV photolysis of diazo(2-thienyl)methane and diazo(3-thienyl)methane in a $N_2$ or Ar ice matrix. Experiments using photons of different wavelengths were carried out, and resulted in a wealth of product molecules containing several interesting structural features and functional groups (Fig. 8), such as: carbenes, sulfur heterocycles, cyclopropenes, alkynes, S=C–C=C bond systems, and S=C=C bond systems. Several of these functionalities may be of astrochemical interest.

# 6  Non-Reactive Charged Particle Radiolysis

In discussing the role of non-reactive charged particles in astrochemistry, it is largely protons and electrons which are considered, although noble gas ions have also been studied. Proton- and electron-irradiation studies are perhaps some of the most explored areas within astrophysical radiochemistry due to their applicability to a wide variety of contexts and scenarios within the space sciences. These include the interaction of interstellar ices with cosmic rays as well as the processing of Solar System ices by the solar wind or magnetospheric ions. With regards to sulfur radiochemistry, early experiments by Moore (1984) found that proton-irradiation of cryogenic $SO_2$ produced several different compounds, and showed that there is a correlation between the colour of the resultant product mixture (which is dependent on its composition) and the temperature of the reaction.

Electron-irradiation of native sulfur in $H_2O$ ice was observed to produce $H_2SO_4$ (Carlson et al. 2002), mirroring results previously obtained by Johnston and Donaldson (1971) and Della Guardia and Johnston (1980), who had irradiated sulfur grains in liquid $H_2O$. Continued processing results in the radiolysis of $SO_4^{2-}$ to form $SO_2$ (Hochanadel et al. 1955), among other compounds. Moore et al. (2007a) later showed that proton-irradiation of $SO_2$ in $H_2O$ ice also resulted in the formation of $H_2SO_4$ which, when subsequently warmed, yielded a variety of hydrates of the acid.

Extensive efforts have been made to understand proton- and electron-irradiation of $SO_2$, either as a pure ice or mixed with $H_2O$, due to the fact that $SO_2$ has been recognised (along with OCS) as a dominant sulfur-bearing molecule in interstellar ices (Maity and Kaiser 2013). Seminal work by Moore et al. (2007b) investigated the 800 keV proton-irradiation of pure $SO_2$ ice and $SO_2:H_2O$ mixed ices of varying compositions at 86-132 K. When the pure $SO_2$ ice was irradiated, $SO_3$ was observed just like in the photolysis experiments of Schriver-Mazzuoli et al. (2003), but this was not the case when a mixed $SO_2:H_2O$ ice underwent radiolysis. Instead, other molecules such as $HSO_3^-$, $HSO_4^-$, $SO_4^{2-}$ (possibly from an acid molecule), and $H_2O_2$ were observed. These results are somewhat analogous to those obtained during the thermal processing of this mixed ice (Loeffler and Hudson 2010; 2013; 2015; Kaňuchová et al. 2017).

An interesting interpretation of these results obtained by Moore et al. (2007b) is the fact that proton implantation in pure $SO_2$ ice does not result in the formation of S–H bonds. Garozzo et al. (2008) confirmed these results by implanting 30 keV protons at 16 K and 50 keV protons at 80 K in pure $SO_2$ ice. The main products of such radiolysis were observed to be $SO_3$ (both as a single molecule and in polymeric form) and $O_3$. Their study also involved the irradiation of $SO_2$ ice by 30 keV $He^+$ ions at 16 K, with similar results being reported.

By showing that no S–H bonds are formed in this way, these studies challenged a previous hypothesis which was made by Voegele et al. (2004), who suggested that the proton-irradiation of $SO_2$ may be a viable method of synthesising $H_2SO_3$ in a manner analogous to the proton-irradiation of $CO_2$ forming $H_2CO_3$ (Brucato et al. 1997). A good summary of these results may also be found in the work of Strazzulla (2011).

Work by Ferrante et al. (2008) and Garozzo et al. (2010) elucidated OCS as the main product of the proton-irradiation of ices containing both sulfur ($SO_2$ or $H_2S$) and carbon (CO or $CO_2$) source molecules. Several scenarios were considered, including water-dominated and water-free ices as well as mixed and layered structures. The overall yield of OCS is determined by the mixing ratio of the ice, as well as the nature of the sulfur and carbon source molecules, with $H_2S$ and CO being the most amenable to OCS formation.

The mechanism of formation of OCS is believed to involve fragmentation of parent source molecules, after which sulfur atoms combine with CO directly (Ferrante et al. 2008). An interesting result of these investigations is the fact that, although OCS forms readily, it is destroyed fairly easily upon prolonged irradiation. It should also be noted that, during the radiolysis of $CO:H_2S$ mixed ices, $CS_2$ is also a relatively abundant product (Garozzo et al. 2010).

The irradiation of $CS_2:O_2$ mixed ices with high-energy electrons has also been studied. Maity and Kaiser (2013) found that, at 12 K, this radiolysis readily converts $CS_2$ to OCS much in the same way as the irradiation of $CS_2$ mixed with $H_2O$, $CO_2$, or $CH_3OH$. Other sulfur-bearing products were also observed, including $SO_2$ and $SO_3$.

Loeffler et al. (2011) performed radiolysis experiments in which $H_2SO_4$, $H_2SO_4 \bullet H_2O$, and $H_2SO_4 \bullet 4H_2O$ were irradiated with 0.8 MeV protons. Such experiments are of importance given that sulfuric acid hydrates are known products of the radiolysis of $SO_2:H_2O$ mixed ices and are distributed widely on Europa (Carlson et al. 1999; 2002; 2005). The main irradiation products were $SO_2$, $S_2O_3$, $H_3O^+$, $HSO_4^-$, and $SO_4^{2-}$. An interesting result of this study, however, was the observed radiolytic stability of the monohydrate acid as compared to that of the pure acid.

Furthermore, destruction of the tetrahydrate acid was noted to be strongly correlated to temperature: increased losses at lower temperatures occurred due to a combination of radiolysis and amorphisation which resulted in a change in the number of water molecules associated with an acid molecule. This poses an interesting result in the context of Europa and the other Galilean satellites, where it is hypothesised that the monohydrate acid is stable over geological time, while the tetrahydrate will only be stable in warmer regions (Loeffler et al. 2011).

The irradiation of $H_2S$ has also been investigated, perhaps as a result of the abundance of this compound in cometary materials (Rodgers and Charnley 2006). Irradiation of $H_2S$ poses a new challenge due to the fact that it sublimates at a relatively low temperature of 86 K at pressures of interstellar relevance. Experiments by Moore et al. (2007b) investigated the irradiation of $H_2S$ as a pure ice and in a mixture with $H_2O$ by 0.8 MeV protons at temperatures between 86-132 K. Irradiation of the pure ice resulted in the formation of $H_2S_2$. This result is of some consequence in the context of astrobiology, as it reveals a fairly facile method of disulfide bond (S–S) formation. Such bonds are important in several protein structures. When $H_2O:H_2S$ mixed ices were irradiated $H_2S_2$ was still observed, but in lower abundances. This is primarily due to the fact that $SO_2$ is also a product of this radiolysis and so competes for sulfur atoms (Moore et al. 2007b).

Electron-irradiation of complex ice mixtures containing $H_2S$ have also been performed recently. Mahjoub et al. (2016; 2017) irradiated a mixture of $H_2S:CH_3OH:NH_3:H_2O$ (in an ice of compositional ratio 7:35:17:41) at 50 K with 10 keV electrons for 19 hours, and then warmed the ice to 120 K at which temperature it was maintained for another hour, all the while being irradiated. After the completion of electron-irradiation, the ice was then warmed to 300 K. Various sulfur-bearing molecules were detected as products of this combined thermal and radiolytic processing, including: OCS, CS, $CS_2$, SO, $SO_2$, $S_2$, $S_3$, $S_4$, $CH_3SCH_3$, $CH_3S_2CH_3$, $CH_3S(O)OCH_3$ and possibly $SO_3$, $S_2O$, and $H_2CSO$.

The ice composition, processing methodology used, and sulfur chemistry observed in the experiments by Mahjoub et al. (2016; 2017) are of direct relevance to Jovian Trojan asteroids. These bodies display a distinct colour bimodality, which is thought to be the result of combined radiolytic and thermal processing of sulfur compounds, particularly short-chain sulfur allotropes. These allotropes are known to be formed by the radiolysis of native sulfur (among other sulfur-bearing compounds) and are highly coloured, exhibiting strong absorption at long wavelengths in the visible spectrum (Meyer 1976; Brabson et al. 1991).

However, such allotropes are also known to be unstable with regards to thermal processing, and combine to form cyclical geometries at higher temperatures (e.g. $S_8$). As the processing methodology used in these experiments is analogous to the temperature fluxes and irradiation regimes to which Jovian Trojans are subjected as they traverse around the sun (Mahjoub et al. 2016; 2017), these results are suggestive of the fact that the observed colour bimodality of these asteroids is at least partly the result of the thermal processing of sulfur allotropes formed radiolytically through interaction with the solar wind.

An interesting set of studies looked into the irradiation of $H_2O$ ice deposited over a refractory sulfurous residue with 200 keV $He^+$ at 80 K in order to determine whether or not this is a possible source of $SO_2$ (Gomis and Strazzulla 2008; Strazzulla et al. 2009). The residue was obtained via the prior irradiation of $SO_2$ ice at 16 K, and was used as an approximation for sulfur-bearing solid materials in astrophysical environments, such as on the surfaces of the Galilean satellites. The studies did not find any evidence of efficient $SO_2$ production, and thus concluded that the radiolysis of mixtures of $H_2O$ ice and refractory sulfurous materials cannot be the primary formation mechanism of the $SO_2$ detected at the surface of the Galilean moons.

Electron-irradiation studies of ionic solids, which may be representative of mineral assemblages at the surfaces of planets and moons, have also been conducted. Sasaki et al. (1978) irradiated $Li_2SO_4$ with 0.3-1.6 keV electrons and detected $Li_2SO_3$, $Li_2S$, $Li_2O$, and elemental sulfur as products. Prolonged irradiation showed that the final radiolysis products were $Li_2S$ and $Li_2O$.

Johnson (2001) considered the irradiation of hydrated $Na_2SO_4$ and $MgSO_4$, both minerals that have been suggested to be present at the geologically younger surfaces of Europa as a result of tidal flexing or volcanism (Kargel 1991; McCord et al. 1998b; 1998c). Ion bombardment of hydrated $MgSO_4$ is thought to produce MgO, MgS, and $Mg(OH)_2$, as well as $O_2$ and $SO_2$ (Sieveka and Johnson 1985; Johnson et al. 1998; Johnson 2001). In the case of hydrated $Na_2SO_4$, irradiation is thought to represent a facile method of sodium loss (Benninghoven 1969; Wiens et al. 1997), and indeed some atomic sodium is a component of the tenuous atmosphere on Europa (Brown and Hill 1996). The net products of irradiation are thought to be NaOH, $Na_2O_2$, $SO_2$, and $H_2SO_4$ (Johnson et al. 1998; Johnson 2001).

Irradiation of these hydrated sulfates also represents a potential formation mechanism for $H_2SO_4$ and $SO_2$ on the surfaces of the icy Galilean satellites. This is of significant consequence because, as will be described in Section 7, there is currently some debate as to whether the latter is formed by magnetospheric sulfur ion implantation of from sulfur-bearing compounds already present on the moons (possibly even a non-radiolytic formation mechanism). However, reports of laboratory irradiations of ionic solids and minerals remain sparse, and so there is a need for more of these experiments to be performed.

# 7  Sulfur Ion Bombardment and Implantation in Ice Analogues

Sulfur astrochemistry and molecular astrophysics plays an important role in understanding the chemistry and surface processes of the icy Galilean satellites of Jupiter. Sulfur-bearing ices on these satellites were detected by the International Ultraviolet Explorer (IUE) mission (Lane et al. 1981) and since the reporting of these early findings, much work has been devoted to understanding the origin and chemistry of these ices.

Such work led to the proposed existence of a radiolytic sulfur cycle (Carlson et al. 1999; 2002; 2005) which includes chemical alteration of the surface by energetic ions and electrons from Jovian magnetospheric plasma (Fig. 9). Laboratory experiments have shown that $H_2SO_4$ and its hydrates may be produced through energetic particle bombardment of mixed ices at the surface of the satellites (Strazzulla et al. 2007; 2009). Indeed, maps of acid production rate show a clear *bulls-eye pattern* which would be expected as a result of Iogenic sulfur ion bombardment (Carlson et al. 2005; 2009). Other sulfur-bearing compounds are also thought to arise from the radiolytic sulfur cycle, with Hendrix et al. (2011) finding a strong correlation between surface $SO_2$ abundance and sulfur ion implantation. A review of the radiolytic sulfur cycle may be found in Dalton et al. (2013).

Because such chemical complexity arises from sulfur ion impacts and implantations, many experiments have been undertaken to further understand the products of the implantation of sulfur ions of various energies and charge states, as well as the mechanisms by which they form. The simplest (and perhaps most investigated) experiments involve sulfur ion implantation in $H_2O$ ice. Such experiments have not found much in the way of evidence to support the formation of $H_2S$ or $SO_2$ from these impacts, despite the findings of Hendrix et al. (2011). Instead, these implantations have largely resulted in the formation of $H_2SO_4$ and its hydrates.

Irradiation of $H_2O$ ice with 200 keV $S^+$ ions at 80 K resulted in the formation of $H_2O_2$ as well as hydrated $H_2SO_4$ (Strazzulla 2011). This acid hydrate was produced with a very high yield of ~0.65 molecules per ion (Strazzulla et al. 2007). These results lead to a new question with regards to the $H_2SO_4$ hydrates present on the surfaces of the icy Galilean satellites (especially Europa): are these acid hydrates the result of Iogenic ion implantation into $H_2O$ ice (i.e. an exogenic sulfur source), or are they the result of some other chemical process involving sulfur compounds native to the icy moon (i.e. and endogenic sulfur source), which forms $SO_3$ which in turn rapidly produces the observed acid hydrates?

Discriminating between exogenic and endogenic sulfur sources is possible by observing the spatial distribution of the observed $H_2SO_4$ hydrates. On Europa, the global distribution of such hydrates is such that there is an enhancement on the trailing side, which is suggestive of an ion implantation (exogenic) source. It is interesting to note that $SO_2$ ice also displays a similar distribution. However, since the concentration of the acid hydrate at the surface is much greater than that of $SO_2$ and because calculations have shown that radiolysis can produce the observed amount of acid hydrate during ~$10^4$ years, it is likely that the overall $H_2SO_4$ hydrate sulfur source on Europa is an exogenic one (Strazzulla 2011).

Further investigations into the irradiation of $H_2O$ ice at 80 K were made using multiply charged $S^{n+}$ ions ($n$ = 7, 9, 11) over an energy range of 35-176 keV (Ding et al. 2013). Results showed that the dominant products formed were $H_2SO_4$ as a pure molecule, as well as in the monohydrate and tetrahydrate forms. The nature of the products was noted to be independent of the charge state of the projectile ion.

This is consistent with the fact that such projectiles abstract electrons from the target surface upon approaching within several tens of angstroms. This so-called *resonant neutralisation* process allows for the formation of transient, neutral *hollow atoms*, in which inner electron shells are not occupied (Arnau et al. 1997; Winter and Aumayr 1999). Electron abstraction leaves a positively charged region at the surface of the solid ice which can explosively expand leading to so-called *potential sputtering* and the formation of *surface hillocks* (Wilhelm et al. 2015). Decay of the transient hollow atoms can occur via radiative electron cascade of Auger electron emission, though time constraints mean that it is often the case where only a few electrons are able to de-excite to the inner shells before impact with the target ice (Herrmann et al. 1994). Once the projectile ion enters the bulk ice, it achieves an effective charge state dependent only upon its impact velocity within a single monolayer, thus losing memory of the original, incoming charge state (Herrmann et al. 1994; Ding et al. 2013).

The yields of the acid products formed were dependent upon the energies of the incoming sulfur ions, with higher energies being responsible for higher yields. Interestingly, no $SO_2$ or $H_2S$ was detected during post-irradiative spectroscopic analysis. Hence, the results of Ding et al. (2013) confirm and extend those obtained by Strazzulla et al. (2007). Ding et al. (2013) also state that their results, combined with the fact that there is a clear correlation between $H_2SO_4$ hydrate concentration and magnetospheric sulfur ion flux, support an exogenic source for sulfur in these molecules on Europa.

The implantation of multiply-charged sulfur ions in other ices has also been studied. It has been shown that the 176 keV $S^{11+}$ ion implantation in CO at 15 K results in the formation of $SO_2$ and OCS, while the 90 keV $S^{9+}$ ion implantation in $CO_2$ ice covered with a thin layer of $H_2O$ ice at the same temperature yields $SO_2$ and $CS_2$ (Lv et al. 2014a). Given the known presence of $CO_2$ on Europa, Lv et al. (2014a) calculated that a time-scale of ~$10^4$ years is required to produce the amount of $SO_2$ present at the surface via sulfur ion bombardment.

However, this calculation runs on the assumption that this yield calculated at 15 K is applicable to temperatures more relevant to the icy surface of Europa (60-100 K). Lv et al. (2014a) also showed that the radiolysis of the CO ice resulted in the formation of a large number of carbon-based chains, while chemical reactions and mixing at the interface of the $CO_2$ and $H_2O$ ices caused the formation of $H_2CO_3$.

Follow-up studies by Boduch et al. (2016), who made use of UV spectroscopy rather than IR spectroscopy as a means of product identification, showed that, when pure $CO_2$, $O_2$, and $H_2O$ ices, as well as mixed $CO_2$ (or $O_2$):$H_2O$ ices were irradiated with 144 keV $S^{9+}$ ions at 16 K, no $SO_2$ was detected among the radiolysis products, in contrast to the results reported by Lv et al. (2014a). Boduch et al. (2016) rationalised that the ion fluence used in their experiments may have meant that the amount of $SO_2$ produced (if any) was less than the detection limit of their spectroscopic instruments.

Additionally, the UV features of $SO_2$ would have been hidden by the strong bands of $HSO_3^-$ and $SO_3^-$, which were detected. These results, coupled with the fact that no detections of $SO_2$ were made when irradiating $H_2O$ ice deposited above refractory sulfurous materials (Gomis and Strazzulla 2008), led Boduch et al. (2016) to propose the possibility of an endogenic sulfur source for the $SO_2$ observed at the surface of Europa, rather than it being the result of magnetospheric sulfur ion bombardment.

Laboratory irradiations of a binary 1:1 $CO_2$:$NH_3$ ice mixture with 144 keV $S^{9+}$ ions have been performed, considering scenarios in which the projectile ion was implanted and when it travelled through the ice (Lv et al. 2014b). Results showed that in both cases, molecules of astrobiological relevance such as ammonium methanoate and dimeric carbamic acid were formed, as well as simple oxides such as $N_2O$ and CO. Similar qualitative results were obtained when considering the irradiation of a ternary ice mixture containing $NH_3$, $CO_2$, and $H_2O$ (Lv et al. 2014b).

Computational studies have also been used to simulate sulfur ion bombardments of astrophysical ice analogues. Molecular dynamics simulations of a 20 MeV $S^+$ ion impacting a complex multi-component ice whose composition is relevant to the Europan surface revealed that this collision results in a net loss of $SO_2$ molecules (which were initially present in the ice mixture) due to their oxidation to $HSO_3^-$ and $SO_3^-$ (Anders and Urbassek 2019a).

These results therefore parallel the laboratory findings of Boduch et al. (2016) and their sulfur ion radiolysis of simpler ices, as well as those of Kaňuchová et al. (2017) who considered the ion irradiation and thermal processing of $SO_2$:$H_2O$ mixed ices. However, this computational simulation did not consider the implantation of the sulfur ion into the ice (Anders and Urbassek 2019a), and so any chemistry resulting from this implantation which could potentially lead to the formation of novel sulfur-bearing compounds was not considered.

The implantation of a 20 MeV $S^+$ ion into an ice containing $H_2O$, $CO_2$, $NH_3$, and $CH_3OH$ was considered in a separate computational study (Anders and Urbassek 2019b). Results showed that the energy imparted by the projectile ion as it traverses through the ice causes fragmentation of the original molecular constituents, which go on to form a wealth of organic species including methanal, methane, methanoic acid, and methoxymethane. More exotic species, such as cyclopropenone and cyclopropanetrione, were also detected.

An interesting result of this simulation was that the projectile sulfur ion did not react after coming to rest within the ice (Anders and Urbassek 2019b). The authors suggest that the lack of sulfur-bearing product molecules may be due to their use of a low fluence of 1 ion per simulation. Thus, the possibility of $SO_2$ observed at the surface of Europa and Ganymede having an exogenic (Iogenic) sulfur source remains a somewhat open question.

Although complex ice mixtures such as that considered by Anders and Urbassek (2019a) are most easily (and cost-effectively) studied via computational means, experimental attempts have also been made. Ruf et al. (2019) irradiated a mixture of 2:1:1 $H_2O$:$NH_3$:$CH_3OH$ with 105 keV $S^{7+}$ ions at 9 K with the aim of characterising organosulfur molecules formed as a result of the radiolysis.

Overall, they identified over 1,100 organosulfur compounds (12% of all assigned signals) through a combination of IR spectroscopic and mass spectrometric techniques. Though perhaps not directly related to the presence of $SO_2$ on the icy surfaces of the Galilean satellites of Jupiter, this finding suggests that sulfur ion implantation could be an impetus for a rich organic chemistry which is significant to several astrophysical contexts, and is undoubtedly of interest from the perspectives of astrobiology and prebiotic chemistry.

# 8  Future Directions

The work reviewed in Sections 2-7 provides a basis for future investigations, and there are several worthwhile and interesting routes that these investigations may follow. For instance, further investigations into the low-temperature IR and UV absorption spectra of relevant molecules will reveal important diagnostic features, and thus aid with their identification in interstellar and circumstellar media. This is perhaps most relevant now that the planned launch date for the JWST mission is approaching (Gardner et al. 2006), as the data collected by this telescope will no doubt increase our repository of known astrochemical molecules.

Laboratory experiments at purpose-built facilities (such as the new Ice Chamber for Astrophysics-Astrochemistry at the ATOMKI Institute for Nuclear Research in Debrecen, Hungary) will also be useful in addressing problems in astrophysical chemistry. The paucity of empirical sulfur neutral-neutral surface reaction experiments, for instance, is one which should be tackled. As previously discussed in Section 3, neutral-neutral reactions are most significant in the context of dense molecular clouds into which UV photons or cosmic rays cannot penetrate to induce chemistry. Several molecular species have been detected in the cores of such environments, and such molecules are most likely formed via surface reactions in ice grains, with a lesser amount formed through gas phase chemistry. Sulfur is likely an important factor here, but is not often considered in laboratory investigations due to the fact that it is a known pollutant in ultra-high vacuum systems. Thus, dedicated laboratory experiments are required to fill this gap in knowledge since the results of such investigations would contribute greatly to our understanding of sulfur depletion in dense interstellar clouds.

Further experiments in solid phase thermal chemistry and photochemistry should also be performed. Such reactions are near ubiquitous in space environments and can occur in any region which is warm enough or is subject to sufficient photon irradiation. Future experiments could therefore reveal much with regards to the sulfur chemistry occurring within diffuse interstellar media and on the surfaces of icy worlds, especially when combined with other processing types such as electron-irradiation. In the case of thermal processing, several interesting prebiotic molecules could be produced via relatively simple chemical steps, such as nucleophilic or electrophilic additions and substitutions. In spite of this, solid phase sulfur thermal chemistry remains poorly characterised.

Performing further radiochemical experiments would also be a worthwhile endeavour since, as explained in Sections 6 and 7, ion bombardment of and implantation in ices occurs in several astrophysical environments, including the diffuse interstellar medium, on planetary and lunar surfaces, and in asteroids and comets. One question which should be addressed is whether the $SO_2$ frosts observed on the surfaces of the Galilean moons, particularly Europa, has an endogenic or exogenic sulfur source.

As it stands, there is contrasting evidence as to whether sulfur ion implantation in $CO_2$ ice can even form $SO_2$ (Lv et al. 2014a; Boduch et al. 2016), with those studies arguing in favour using too low a temperature regime to be directly applicable to the Europan surface. Dedicated experiments could yield more information as to whether $SO_2$ is a product of sulfur ion implantation in CO or $CO_2$ ice, and also whether it is produced with sufficient yields to explain the observed quantities of $SO_2$ on Europa. Such experiments should be systematic in nature, observing the results of sulfur ion implantations of various projectile charge states and energies in CO and $CO_2$ ices at a range of temperatures.

As discussed in Section 1, there is a severe lack of systematic investigations in the field of astrochemistry, with many studies simply reporting the results of a very specific set of reaction conditions (particular temperatures, projectile natures and energy states, ice morphology and composition, etc.). Although such studies surely contribute to our knowledge of chemical reactivity in extra-terrestrial environments, it is difficult to gauge the applicability of their results to different astrophysical settings.

For instance, the results of a study conducted at 20 K may be relevant to ices in the interstellar medium, but not so to icy planetary and lunar surfaces where temperatures are higher. Another example is ice thickness: experiments which look into the astrochemical processing of thin ices may not be applicable to icy worlds, where ices are kilometres thick. Thus, radiolysis of thin ices would likely allow for the projectile to travel through the ice, whereas implantation would certainly occur on an icy world. Thus, for the reasons outlined above, there is a genuine need for systematic investigations of sulfur chemistry in space and planetary environments. Furthermore, such studies will be of enormous benefit is assessing the data collected by upcoming space missions, including JUICE and JWST.

Another potential avenue of investigation involves aligning more closely the fields of astrochemistry and cosmochemistry. Isotope studies are important in the space sciences and underpin many meteoritic and Solar System studies (McSween and Huss 2010). An example of a potential astrochemical study with cosmochemical implications involves the vacuum-UV photo-dissociation of $H_2S$ to yield elemental sulfur. In the gas phase, this is known to be accompanied by sulfur isotope fractionation which is dependent upon the irradiation wavelength used (Chakraborty et al. 2013). However, to the best of the authors' knowledge no analogous studies have been conducted in the solid phase. Thus, a future investigation may well look into any sulfur isotope fractionations occurring during the photolysis of $H_2S$ ice at interstellar temperatures. The radiolysis of mineral assemblies could also be another interesting avenue of investigation.

Finally, we note that that we have not discussed the sputtering of material that often occurs concomitantly during radiolysis and photolysis of astrophysical ice analogues (Muntean et al. 2015; 2016). Nevertheless, such a phenomenon is important as it occurs in most laboratory experiments and is thought to contribute to the transient atmospheres on the Galilean moons (Shematovich et al. 2005; Plainaki et al. 2012), and so an effort at furthering our understanding of the topic should be made. Given that sputtering is a well-established subject within the field of astrophysical chemistry and molecular astrophysics, we direct the interested reader to the reviews by Baragiola et al. (2003) and Famá et al. (2008).

# 9     Conclusions

This review has highlighted the major and recent findings by laboratory and (to a lesser extent) computational studies in condensed phase sulfur astrochemistry. Potential future directions have also been discussed. Although perhaps not a complete survey of the field, several important points have been communicated, in particular:

- Neutral-neutral reactions relevant to extra-terrestrial sulfur chemistry have yet to be rigorously explored and are sorely lacking in experimental data compared to analogous gas phase studies.

- The cryogenic thermal chemistry of mixed ices containing sulfur-bearing molecules is potentially rich and has not yet been fully explored.

- Photolysis of pure $SO_2$ ice yields $SO_3$, while that of pure $H_2S$ yields products such as HS, $HS_2$, $H_2S_2$, and $S_2$.

- Photolysis of mixed $SO_2$:$H_2O$ ice results in the formation of $H_2SO_4$ and related sulfur oxyanions.

- When mixed with CO or $CO_2$, photolysis of $H_2S$ yields OCS. $CS_2$ is also produced if the carbon-bearing molecule is CO, while $SO_2$ is a by-product when it is $CO_2$.

- Proton-irradiation of $SO_2$ has not been shown to form S–H bonds, rather $SO_3$ is formed in both molecular and polymeric form.

- Proton-irradiation of mixed $SO_2$:$H_2O$ ices results in the formation of $H_2SO_4$ acid hydrates and related molecules and fragments.

- Proton-irradiation of ices containing both sulfur and carbon source molecules allows for the formation of OCS and $CS_2$.

- $SO_2$, $HSO_4^-$, and $SO_4^{2-}$ are the major radiolysis products of the proton-irradiation of $H_2SO_4$ and its hydrates.

- The main product of sulfur ion radiolysis of $H_2O$ ice is $H_2SO_4$ and its hydrates. The charge state of the projectile ion does not influence the resultant chemistry, but the projectile energy is correlated with the yield of the acid product.

- There is evidence for an exogenic (Iogenic) sulfur source for $H_2SO_4$ on the surface of Europa and possibly also the other icy Galilean satellites.

- There is still a need for further investigation into the sulfur source for $SO_2$ on the surface of Europa (i.e. whether it is endogenic or exogenic).

- It is important that future studies make use of a systematic experimental design in which multiple factors (e.g. different temperatures, ice morphologies, projectile ion charge states and energies, etc.) are considered, as this would allow experimental results to be applied to several astrophysical environments.

It is also important to once again draw attention to the fact that sulfur astrochemistry is by no means limited to the condensed phase, and that gas phase chemistry is likely to be a major contributor in settings within and beyond the Solar System. Furthermore, the space chemistry of sulfur may also be investigated mineralogically and isotopically through cosmochemical experiments, and such experiments may also have implications for geochemical processes on Earth.

## 10    Summary of Review

Table 2 summarises in brief the laboratory solid phase sulfur astrochemistry investigations reviewed in this paper. Although full details are found in text, this summary may be used as a quick reference guide.

**Table 2:** Quick reference guide for sulfur astrochemical experiments reviewed in this work.

| Type | Experiment | Reference |
|---|---|---|
| N | Atomic oxygen reacted with $SO_2$ | Schriver-Mazzuoli et al. (2003) |
| N | Atomic oxygen reacted with $CS_2$ | Ward et al. (2012) |
| N | Atomic carbon reacted with OCS, $CS_2$, $SO_2$, and $H_2S$ | Deeyamulla and Husain (2006) |
| N | Atomic sulfur reacted with $S_n$ ($n$ = 2, 3, 4, 5, 6, 7) | Barnes et al. (1974) |
| N | Atomic sulfur reacted with $H_2S$ | Jiménez-Escobar and Muñoz-Caro (2011) |
| N | Atomic hydrogen reacted with HS and $HS_2$ | Jiménez-Escobar and Muñoz-Caro (2011) |
| T | $SO_2$ reacted with $H_2O$ | Loeffler and Hudson (2010); Bang et al. (2017); Kaňuchová et al. (2017) |
| T | Reaction between $SO_2$, $H_2O$, and $H_2O_2$ | Loeffler and Hudson (2013) |
| T | $SO_2$ reacted with $H_2O_2$ | Loeffler and Hudson (2015) |
| T | Reaction between $SO_2$, $H_2O$, and $O_3$ | Loeffler and Hudson (2017) |
| T | Methylamine reacted with OCS | Mahjoub and Hodyss (2018) |
| P | UV photolysis of $H_2S$ | Harrison et al. (1988); Liu et al. (1999); Cook et |

| Type | Experiment | Reference |
|---|---|---|
| | | al. (2001); Zhou et al. (2020) |
| P | UV photolysis of $H_2S:H_2O$ | Jiménez-Escobar and Muñoz-Caro (2011) |
| P | UV photolysis of $H_2S:CO$ | Jiménez-Escobar et al. (2014); Chen et al. (2015) |
| P | UV photolysis of $H_2S:CO_2$ | Chen et al. (2015) |
| P | UV photolysis of $H_2S:CH_3OH$ | Jiménez-Escobar et al. (2014) |
| P | UV photolysis of $SO_2$ | Sodeau and Lee (1980); Schriver-Mazzuoli et al. (2003) |
| P | UV photolysis of $SO_2:H_2O$ | Schriver-Mazzuoli et al. (2003); Hodyss et al. (2019) |
| P | UV photolysis of $SO_2:O_2$ | Sodeau and Lee (1980) |
| P | UV photolysis of OCS | Dixon-Warren et al. (1990); Leggett et al. (1990); Ikeda et al. (2008) |
| P | UV photolysis of $CS_2$ | Ikeda et al. (2008) |
| P | UV photolysis of OCS with $Br_2$, $Cl_2$, ClI, and BrI | Romano et al. (2001); Tobón et al. (2006) |
| P | UV photolysis of $CS_2$ with $Cl_2$, $Br_2$, ClBr | Tobón et al. (2007) |
| P | UV photolysis of $HSCH_2CN$ | Zapała et al. (2019) |
| P | UV photolysis of $C_4H_3SCHN_2$ structural isomers | Pharr et al. (2012) |
| P | X-ray photolysis of $SO_2$ | de Souza Bonfim et al. (2017) |
| P | X-ray photolysis of $H_2O:CO_2:NH_3:SO_2$ | Pilling and Bergantini (2015) |
| R | $H^+$ radiolysis of $H_2S$, $SO_2$, and OCS | Moore (1984); Moore et al. (2007b); Ferrante et al. (2008); Garozzo et al. (2008) |
| R | $H^+$ radiolysis of $H_2S:H_2O$ | Moore et al. (2007b) |
| R | $H^+$ radiolysis of $H_2S:CO$ and $H_2S:CO_2$ | Ferrante et al. (2008); Garozzo et al. (2010) |
| R | $H^+$ radiolysis of $SO_2:H_2O$ | Moore et al. (2007b) |
| R | $H^+$ radiolysis of $SO_2:CO$ and $SO_2:CO_2$ | Ferrante et al. (2008); Garozzo et al. (2010) |
| R | $H^+$ radiolysis of $H_2SO_4 \cdot nH_2O$ ($n = 0, 1, 4$) | Loeffler et al. (2011) |
| R | $He^+$ radiolysis of $H_2S$ | Strazzulla et al. (2009) |
| R | $He^+$ radiolysis of $SO_2$ | Garozzo et al. (2008); Gomis and Strazzulla (2008) |

| Type | Experiment | Reference |
|---|---|---|
| R | $He^+$ radiolysis of $SO_2$:$H_2O$ | Gomis and Strazzulla (2008); Kaňuchová et al. (2017) |
| R | $Ar^+$ radiolysis of $SO_2$ | Gomis and Strazzulla (2008) |
| R | $Ar^{9+}$ radiolysis of $H_2O$:$O_2$ | Boduch et al. (2016) |
| R | $S^+$ radiolysis of $H_2O$ | Strazzulla et al. (2007) |
| R | $S^{n+}$ radiolysis of $H_2O$, $O_2$, $CO$, $CO_2$, $H_2O$:$O_2$, $H_2O$:$CO_2$, $CO_2$:$NH_3$, $H_2O$:$NH_3$:$CH_3OH$ ($n$ = 7, 9, 11) | Ding et al. (2013); Lv et al. (2014a); Lv et al. (2014b); Boduch et al. (2016); Ruf et al. (2019) |
| R | $e^-$ radiolysis of $H_2O$:$NH_3$:$CH_3OH$ and $H_2O$:$NH_3$:$CH_3OH$:$H_2S$ | Mahjoub et al. (2016); Mahjoub et al. (2017) |
| R | $e^-$ radiolysis of $CS_2$:$O_2$ | Maity and Kaiser (2013) |
| R | $e^-$ radiolysis of $Li_2SO_4$ | Sasaki et al. (1978) |
| R | $\gamma$-ray radiolysis of $S_8$:$H_2O$ | Donaldson and Johnston (1971); Della Guardia and Johnston (1980); Carlson et al. (2002) |
| R | $\gamma$-ray radiolysis of $H_2SO_4$ | Hochanadel et al. (1955) |

*Key: N = neutral-neutral reactions, T = thermal reactions, P = photochemical reactions, R = radiochemical reactions.*

# 11 Declarations

The following declarations are made in line with journal requirements:

## 11.1 Funding

This research has received support from the Europlanet 2024 RI which has received funding from the European Union's Horizon 2020 research innovation programme under Grant Agreement No. 871149.

## 11.2 Conflicts of Interest or Competing Interests

The authors hereby declare that they have no conflicts of interest or competing interests.

## 11.3 Availability of Data and Material

Not applicable.

## 11.4 Code Availability

Not applicable.

**Figures and Figure Captions**

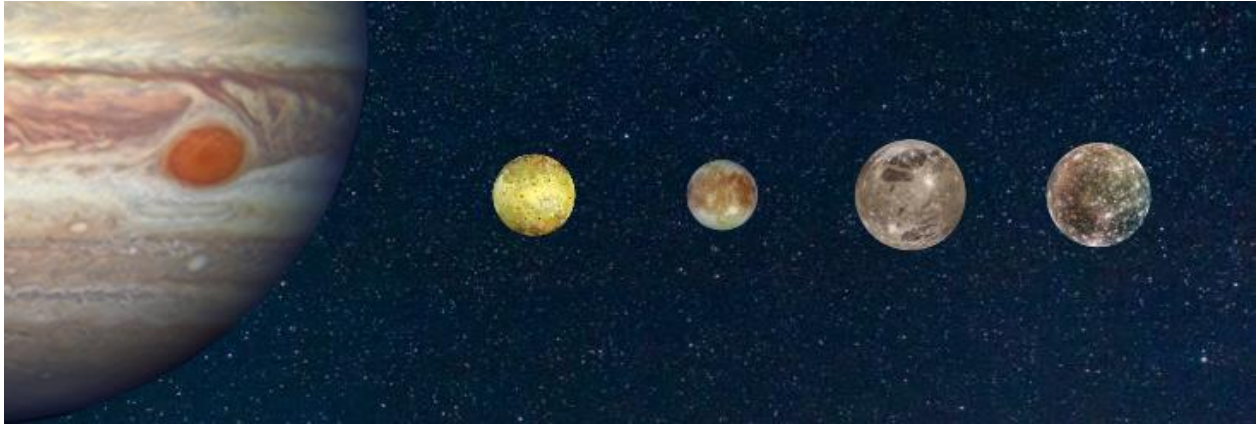

**Fig. 1** The Galilean moon system of Jupiter (relative sizes and orbital distances not to scale). Original image credits to NASA/JPL

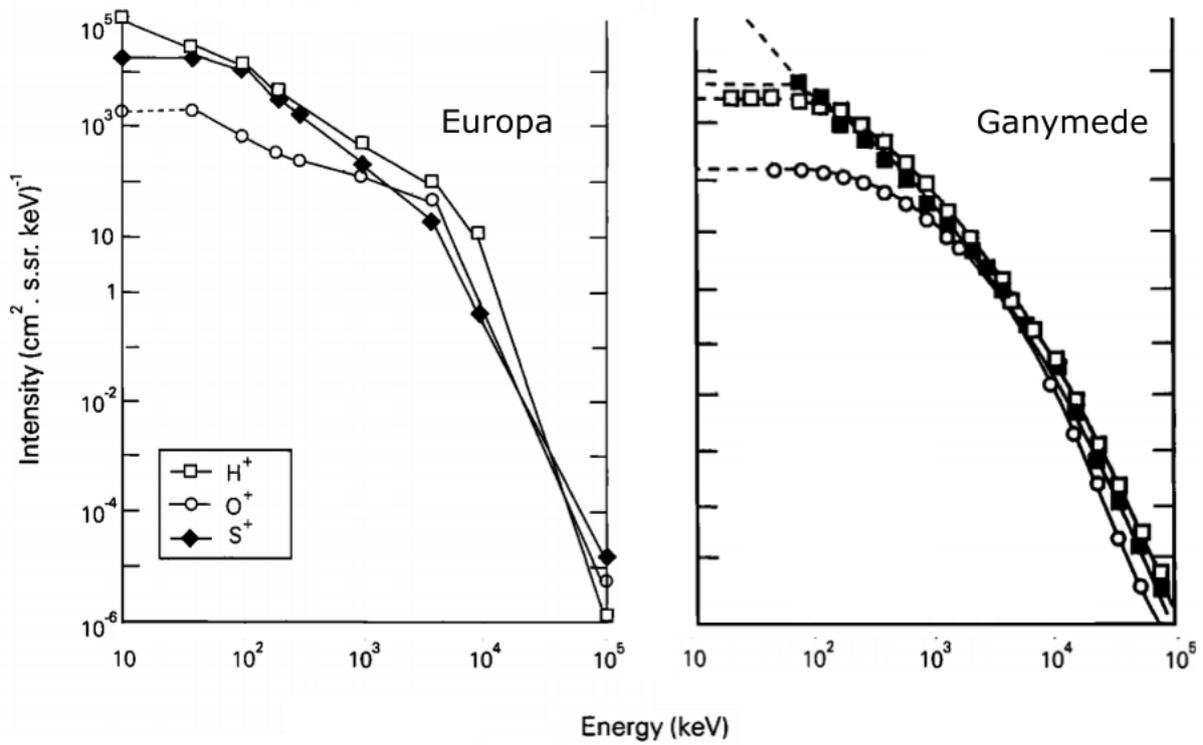

**Fig. 2** Energy profiles of Jovian magnetospheric protons, O$^+$ ions, and S$^+$ ions near the orbits of Europa and Ganymede. Data originally from Ip et al. (1997; 1998). Copyrighted AGU. Reproduced with permission

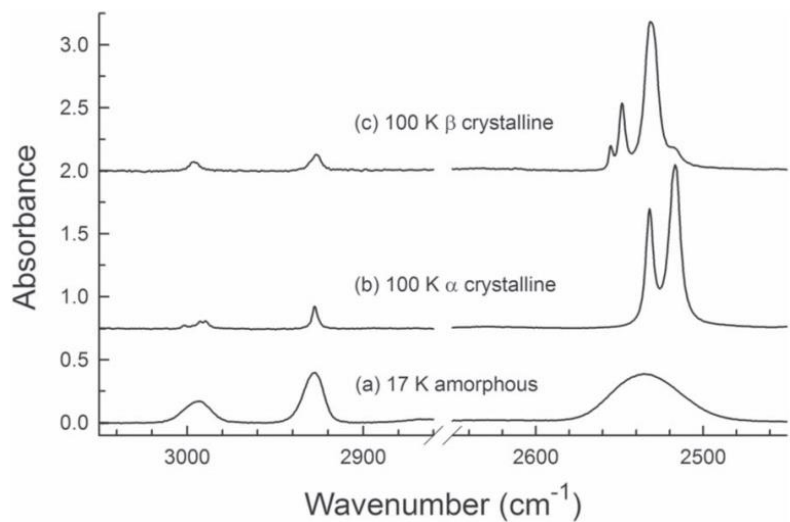

**Fig. 3** IR absorbance spectra for methanethiol: peaks at lower wavenumbers (~2550 cm$^{-1}$) correspond to S–H modes while peaks at higher wavenumbers (~2850-3000 cm$^{-1}$) correspond to C–H stretching modes. Individual spectra shifted vertically for clarity. Data originally from Hudson and Gerakines (2018). Copyrighted AAS. Reproduced with permission

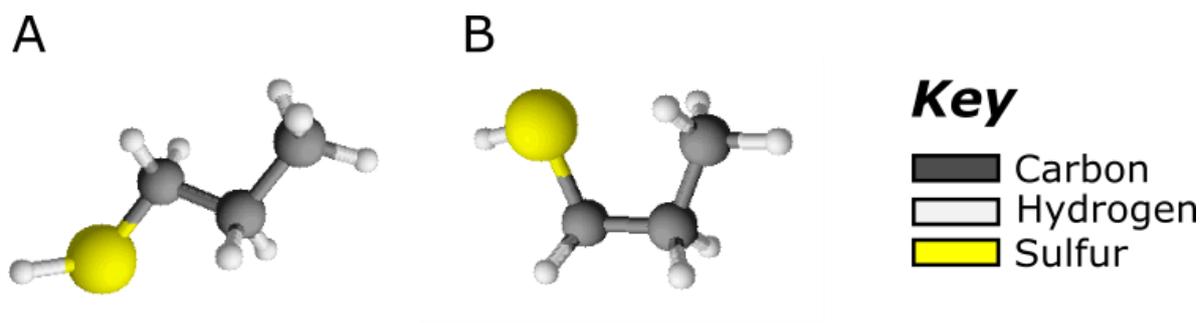

**Fig. 4** Conformational isomerism exhibited using 1-propanethiol as an example. Note that, when looking down the $C_1$–$C_2$ single bond, the thiol and methyl functional groups are on opposite sides (the *anti*-periplanar rotamer) in structure A. In structure B, however, they face each other directly (the *syn*-periplanar rotamer). The other rotamers in this system, the gauche and eclipsed rotamers, are not shown for clarity

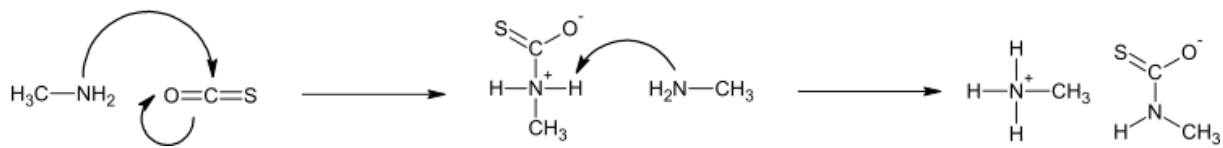

**Fig. 5** Nucleophilic attack occurs during which the lone electron pair on the amine nitrogen bonds with the electron-deficient carbonyl carbon. This causes electron re-distribution from the π-electrons and the development of a formal negative charge on the oxygen atom. Proton abstraction by a second methylamine molecule furnishes the final ionic product

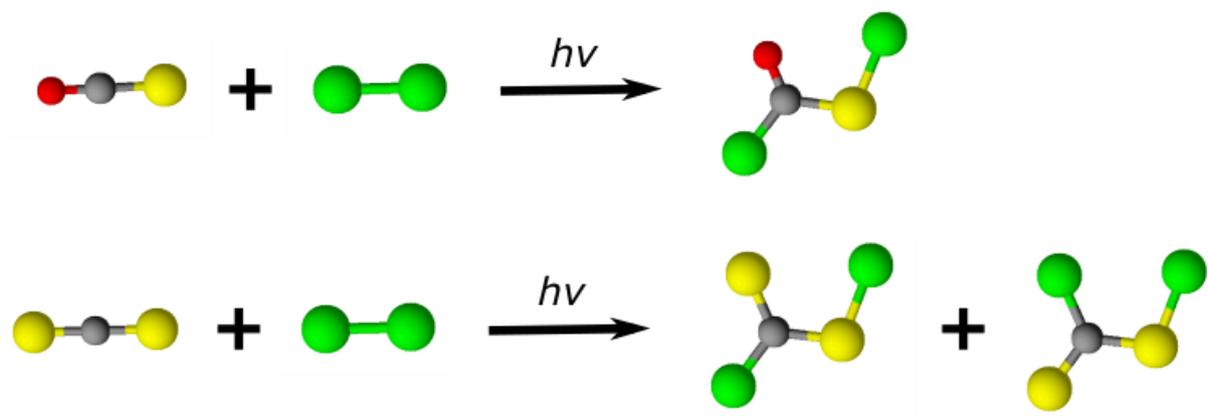

**Fig. 6** The cryogenic solid-phase reaction of OCS with $Cl_2$, $Br_2$, or BrI yields *syn*-halogenocarbonylsulfenyl halides (Romano et al. 2001; Tobón et al. 2006), while the reaction of $CS_2$ with $Cl_2$, $Br_2$, or ClBr under similar conditions yields both *syn*- and *anti*-halogenothiocarbonylsulfenyl halides (Tobón et al. 2007). Note that, in the former reaction, iodine atoms bond directly to the carbonyl group only

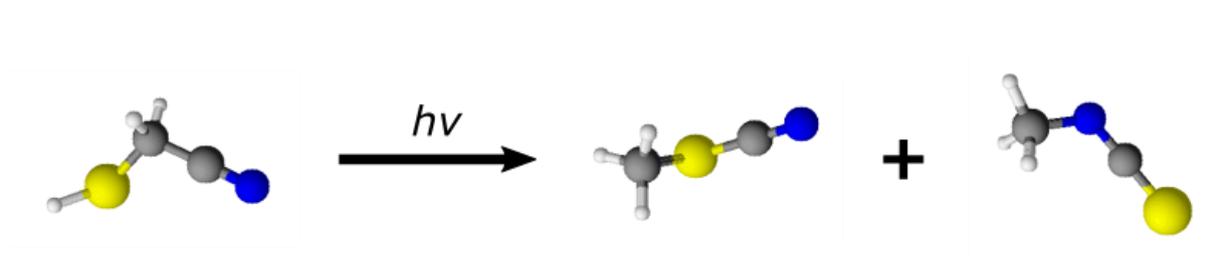

**Fig. 7** Cryogenic photolysis of 2-sulfanylethanenitrile results in the formation of structural isomeric products (Zapała et al. 2019)

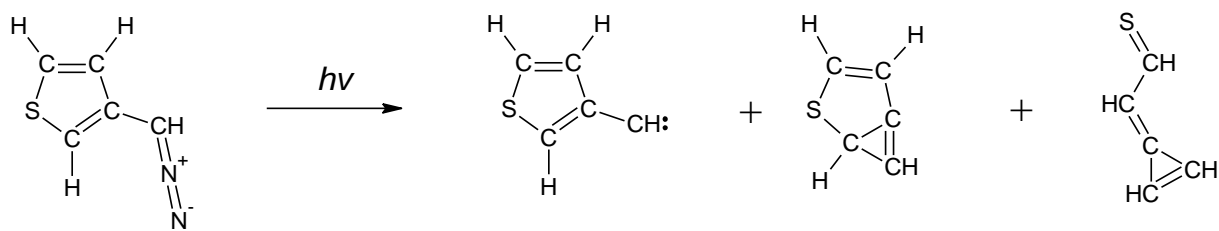

**Fig. 8** The solid phase photolysis of diazo(3-thienyl)methane results in the formation of molecules with functional groups which may be of astrochemical interest (Pharr et al. 2012)

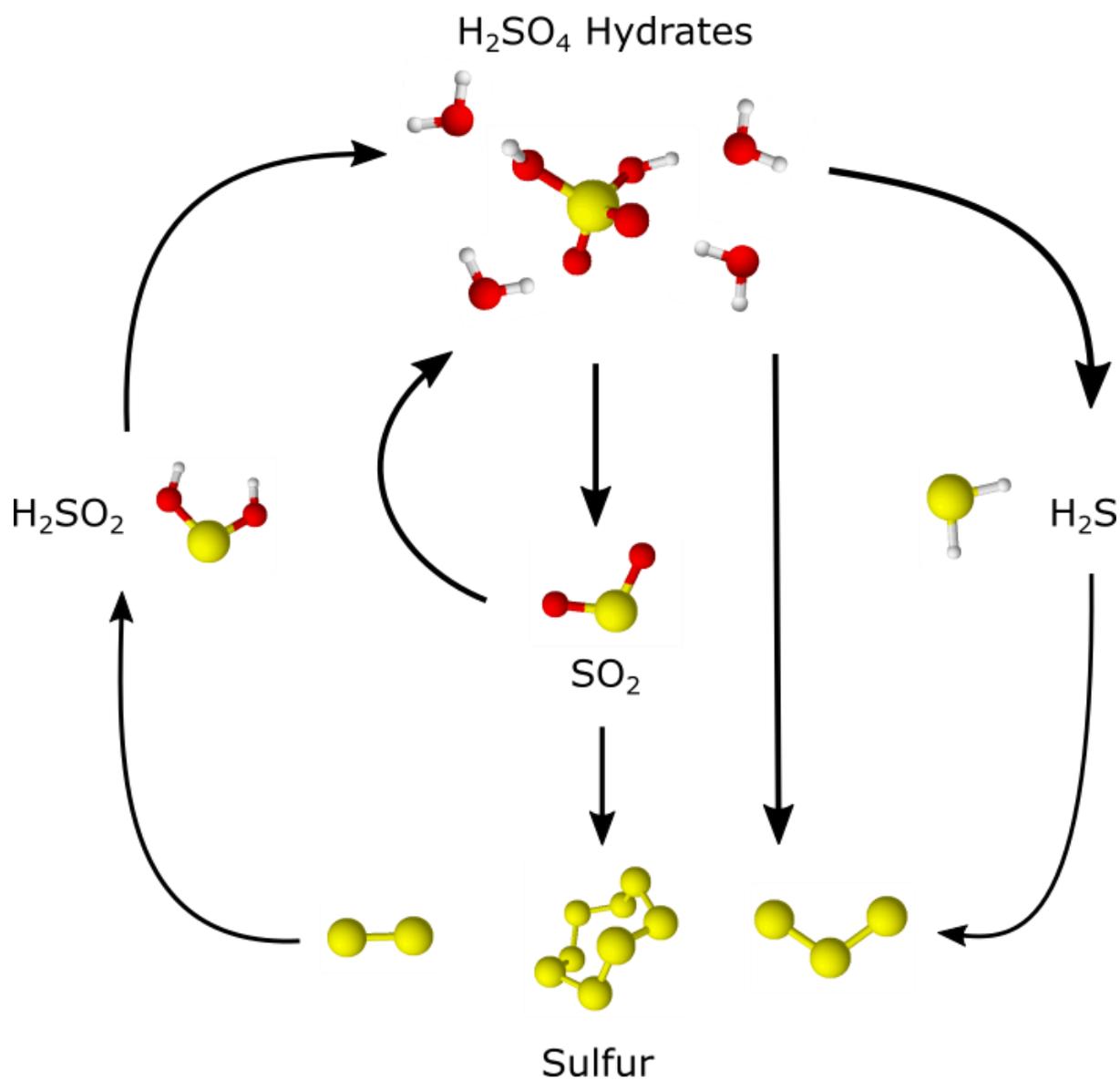

Fig. 9 A qualitative representation of the Europan radiolytic sulfur cycle, wherein arrows indicate radiolysis pathways. The full cycle is completed in ~4000 years. Further details on process rates and species lifetimes may be found in Carlson et al. (2002)